\documentclass[preprint,12pt]{elsarticle}

\usepackage{graphics}

\usepackage{amssymb}
\usepackage{comment}
\usepackage{subfig}

\journal{}

\begin{document}

\begin{frontmatter}



\title{Anger is More Influential Than Joy: Sentiment Correlation in Weibo}


\author{Rui Fan, Jichang Zhao, Yan Chen and Ke Xu\footnote{To whom the correspondence should be addressed: kexu@nlsde.buaa.edu.cn.}}

\address{State Key Laboratory of Software Development Environment, Beihang University, Beijing 100191, P.R.China}

\begin{abstract}
Recent years have witnessed the tremendous growth of the online social media. In China, Weibo, a Twitter-like service, has attracted more than 500 million users in less than four years. Connected by online social ties, different users influence each other emotionally. We find the correlation of anger among users is significantly higher than that of joy, which indicates that angry emotion could spread more quickly and broadly in the network. While the correlation of sadness is surprisingly low and highly fluctuated. Moreover, there is a stronger sentiment correlation between a pair of users if they share more interactions. And users with larger number of friends posses more significant sentiment influence to their neighborhoods. Our findings could provide insights for modeling sentiment influence and propagation in online social networks.

\end{abstract}


\end{frontmatter}


\section{Introduction}
\label{sec:intro}

From the view of conventional social theory, \emph{homophily} leads to connections in social network, as the saying ``Birds of a feather flock together" states~\cite{homophily}. Even in online social network, more and more evidence indicates that the users with similar properties would be connected in the future with high probabilities~\cite{trustties,link_prediction}. It is clear that \emph{homophily} could affect user behavior both online and offline~\cite{happiness_assortative,boost_forum}, while the records in online social network are relatively easier to be tracked and collected. Moreover, the continuous growth of the online social media attracts a vast number of internet users and produces many huge social networks. Twitter\footnote{www.twitter.com}, a microblogging website launched in 2006, has over 200 million active users, with over 140 million microblog posts, known as tweets, being posted everyday. In China, Weibo\footnote{www.weibo.com}, a Twitter-like service launched in 2009, has accumulated more than 500 million registered users in less than four years. Everyday there will be more than 100 million Chinese tweets published. The high-dimension content generated by millons of global users is a ``big data" window~\cite{happiness_correlation} to investigate the online social networks. That is to say, these large-scale online social networks provide an unprecedented opportunity for the study of human behavior.

Beyond typical demographic features such as age, race, hometown, common friends and interest, \emph{homophily} also includes psychological states, like loneliness and happiness~
\cite{homophily,facebook-community,happiness_assortative}. Previous studies also show that the computer-mediated emotional communication is similar to the traditional face-to-face communication, which means there is no evident indication that human communication in online social media is less emotional or less personally~\cite{cmc_emotion}. The tweets posted in online social networks deliver not only the factual information but also the sentiment of the users, which represents their reflections on different social events. Recent study~\cite{happiness_assortative} shows that happiness is assortative in Twitter network and~\cite{happiness_correlation} finds that the average happiness scores are positively correlated between the Twitter users connected by one, two or three social ties. While in these studies, the human emotion is simplified to two classes of positive and negative or just a score of general happiness, neglecting the detailed aspects of human sentiment, especially the negative emotion. Because of oversimplification of the emotion classification, it is hard for the previous literature to disclose the different correlations of different sentiments and then make comparisons. However, the negative emotions, like anger, sadness or disgust, are more applicable in real world scenarios such as abnormal event detection or emergency spread tracking. Figuring out the correlation of these emotions could shed light on why and how the abnormal event begins to spread in the network and then leads to large-scale collective behavior across the entire network. On the other hand, the investigation of how the local structure affects the emotion correlation is not systematically performed yet, while which is essential to studying the mechanism of sentiment influence and contagion.

Aiming at fill these vital gaps, we divide the sentiment of a person into four categories, including \emph{anger}, \emph{joy}, \emph{sadness} and \emph{disgust}, and investigate the emotion correlation between connected users in the interaction network obtained from Weibo. Out of our expectation, it is found that \emph{anger} has a stronger correlation between different users than that of \emph{joy}, while \emph{sadness}'s correlation is trivial. This indicates that \emph{anger} could propagates fast and broadly in the network, which could explain why the real-world events about food security, government bribery or demolition scandal are always the hot trend in internet of China. Moreover, node degree and tie strength both could positively boost the emotion correlation in online social networks. Finally, We make our datasets in this paper public available to the research community.

The rest of this paper is organized as follows. In Section~\ref{sec:relatedworks}, closely related literature would be reviewed, including the methods of sentiment analysis and the difference between our contributions and the previous findings. The data sets employed in this paper and the methods of emotion classification would be introduced in Section~\ref{sec:methods}. We also define the correlation of emotion in this section. Section~\ref{sec:results} reports our findings in detail and our empirical explanations and several real-world case studies would be elucidated in Section~\ref{sec:explanation}. Finally, we give a further discussion in Section~\ref{sec:conclusion} and then conclude this paper briefly.

\section{Related works}
\label{sec:relatedworks}

The content in online social media like Twitter or Weibo is mainly recorded in the form of text. Many approaches have been presented to mine sentiment from these texts in recent years. One of them is the lexicon based method, in which the sentiment of a tweet is determined by counting the number of sentimental words, i.e., positive terms and negative terms. For example, Dodds and Danforth measured the happiness of songs, blogs and presidents~\cite{Dodds_happiness}. They also employed Amazon Mechanical Turk to score over 10000 unique English words on an integer scale from 1 to 9, where 1 represents sad and 9 represents happiness~\cite{Dodds_plosone}. Golder and Macy collected 509 million English tweets from 2.4 million users in Twitter, then measured the positive and negative affects using Linguistic Inquiry and Word Count(LIWC) (http://www.liwc.net). While another one is the machine learning based solution, in which different features are considered to perform the task of classification, including terms, smileys, emoticons and etc. The first step is taken by Pang et al. in ~\cite{pang_acl_2002}, they treat the sentiment classification  of movie reviews  simply as a text categorization problem and investigate several typical classification algorithms. According to the experimental results, machine learning based classifiers outperform the baseline method based on human words list~\cite{twitter_various_techniques,read_acl2005,baseline_method}. 
Different from most work which just categorized the emotion into negative and positive, ~\cite{moodlens} divided the sentiment into four classes, then presented a framework based on emoticons without manually-labelled training tweets and achieved a convincing precision. Because of the ability of multi-emotions classification, we employ this framework in the present paper. 

Each user in the online social network could be a social sensor and the huge amount of tweets convey complicated signals about the users and the real-world events, among which the sentiments are an essential part. Emotion states of the users play a key role in understanding the user behaviors in social networks, whether from an individual or group perspective. In addition, users' mood states are significantly affected by the real-world events~\cite{Bollen_public_mood}.~\cite{Bollen_predict_market} employed the public mood states to predict the stock market and~\cite{moodlens} found the variation of the emotion could be used to detect the abnormal event in real-world, especially the negative sentiment. Individual happiness was measured and several temporal patterns of happiness were revealed in~\cite{Dodds_plosone}. In~\cite{Golder_science}, Golder and Macy collected 509 million English tweets from 2.4 million users in Twitter and disclosed the individual-level diurnal and seasonal mood rhythms in cultures across the globe. The population's mood status was also used to conduct the political forecasting~\cite{distant_politics_ecacl2012}. About the emotion correlation, recent studies~\cite{happiness_assortative,happiness_correlation} show that happiness is assortative in Twitter. While other negative emotions' correlations are not considered in these studies and how the local structure affect the sentiment influence is also not fully investigated. We try to focus more on the difference of correlation between different sentiments and probe deeper into the relation between local structure and emotion correlation. We conjecture that emotion play a significant role in information contagion, especially the negative emotions~\cite{boost_forum}. Because of this, understanding the correlation difference could shed light on the origin of abnormal event propagation in online social media and provide many inspirations for modeling sentiment influence. 

\section{Methods}
\label{sec:methods}

In this section, the methodology of the present paper would be depicted. First, we introduce the collection of the tweets from Weibo and the construction of the interaction network. Then the classifier we employ to mine the sentiment from tweets is reported. Thirdly we define two kinds of emotion correlations for connected users in the interaction network. 

\subsection{Weibo Dataset}
\label{subsec:datasets}

\begin{figure}
\centering
\includegraphics[width=7.5cm]{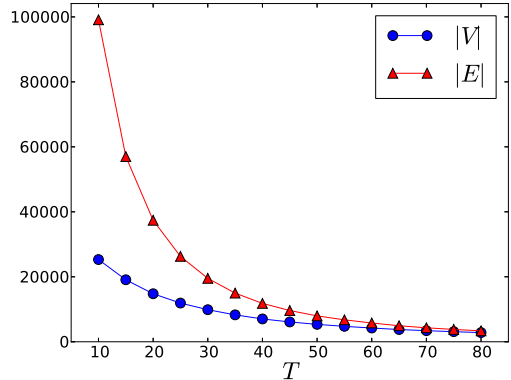}
\caption{The number of nodes or edges varies for different interaction threshold $T.$ In the following part of the present work, we set $T=30$ to extract a large enough network with convincing interaction strength.}
\label{fig:dataset}
\end{figure}

As pointed out in~\cite{happiness_assortative}, the following relationship in Twitter-like social networks does not stand for the social interaction, while if two users reply, retweet or mention each other in their tweets for certain times, the online social tie between them is sufficient to present an alternative means of deriving a conventional social network~\cite{happiness_correlation}. So here we construct an interaction network from the tweets we crawled from Weibo during April 2010 to September 2010, where interaction means the number that two users retweet or mention each other is larger than a threshold $T$. From around 70 million tweets and 200,000 users we crawled, an undirected but weighted graph $G(V,E,T)$ is constructed, in which $V$ is the set of users, $E$ represents the set of interactive links among $V$ and $T$ is the minimum number of interactions on each link. For each link in $E$, its weight is the sum of retweet or mention times between its two ends in the specified time period. Specifically, to exclude occasional users that are not truly involved in the Weibo social network, we only reserve those active users in our interaction network that posted more than one tweet every two days on average over the six months. And to guarantee the validity of users' social interaction, if the number of two users retweet or mention each other is less than $T$, we would omit the connection between them. As shown in Figure~\ref{fig:dataset}, by tuning $T$ we can obtain networks of different scales. Generally we set $T=30$ and then the interaction network $G$ contains 9868 nodes and 19517 links. We also make our entire dataset publicly available\footnote{http://ipv6.nlsde.buaa.edu.cn/zhaojichang/emotionspread.tar.gz}. 

\subsection{Emotion classification}
\label{subsec:emotionclassification}

\begin{figure*}
\centering
\subfloat[An interaction network.]{\includegraphics[width=6cm]{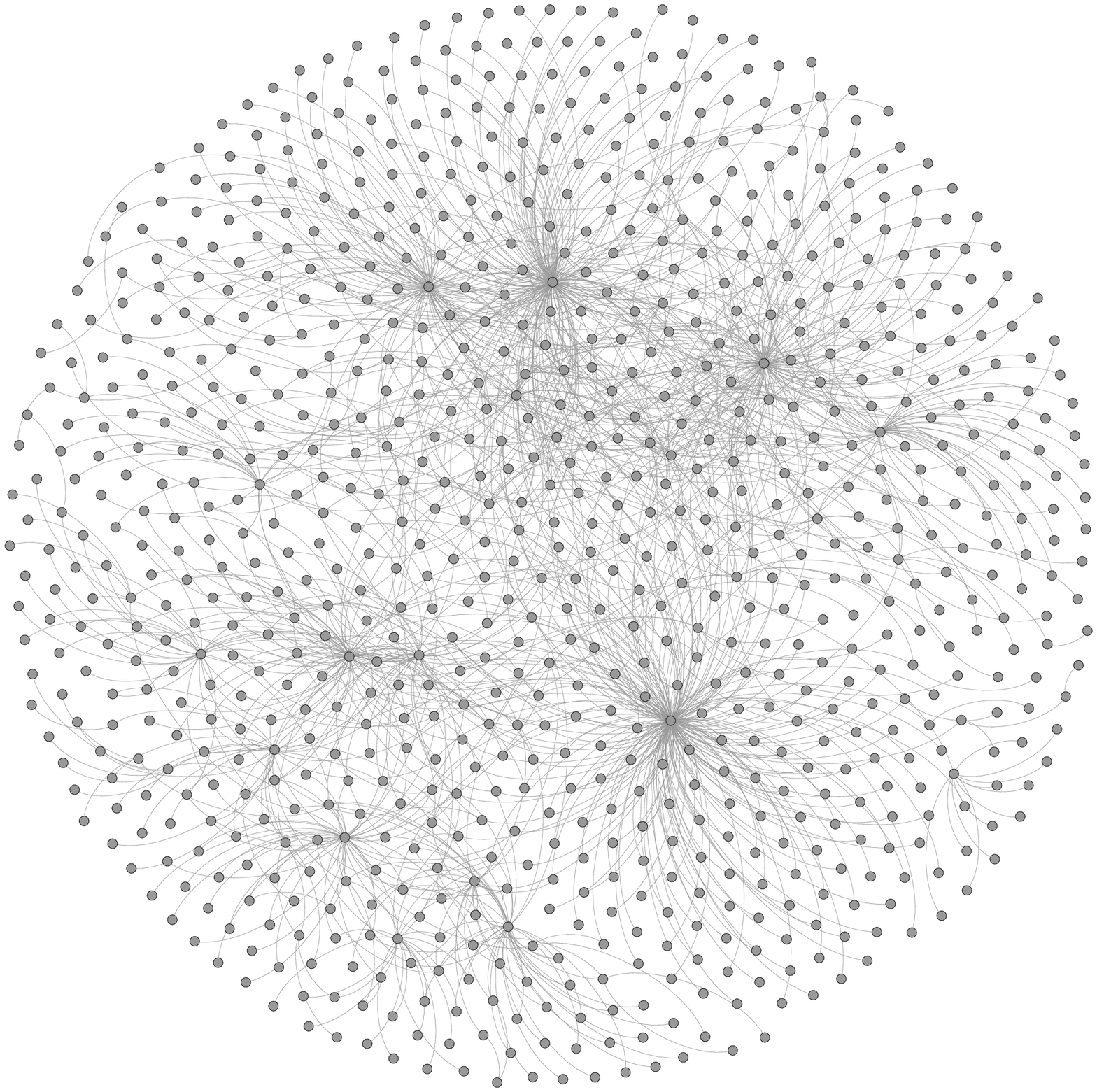}
 \label{fig:topo_init}}
\subfloat[Node colored by emotions.]{\includegraphics[width=6cm]{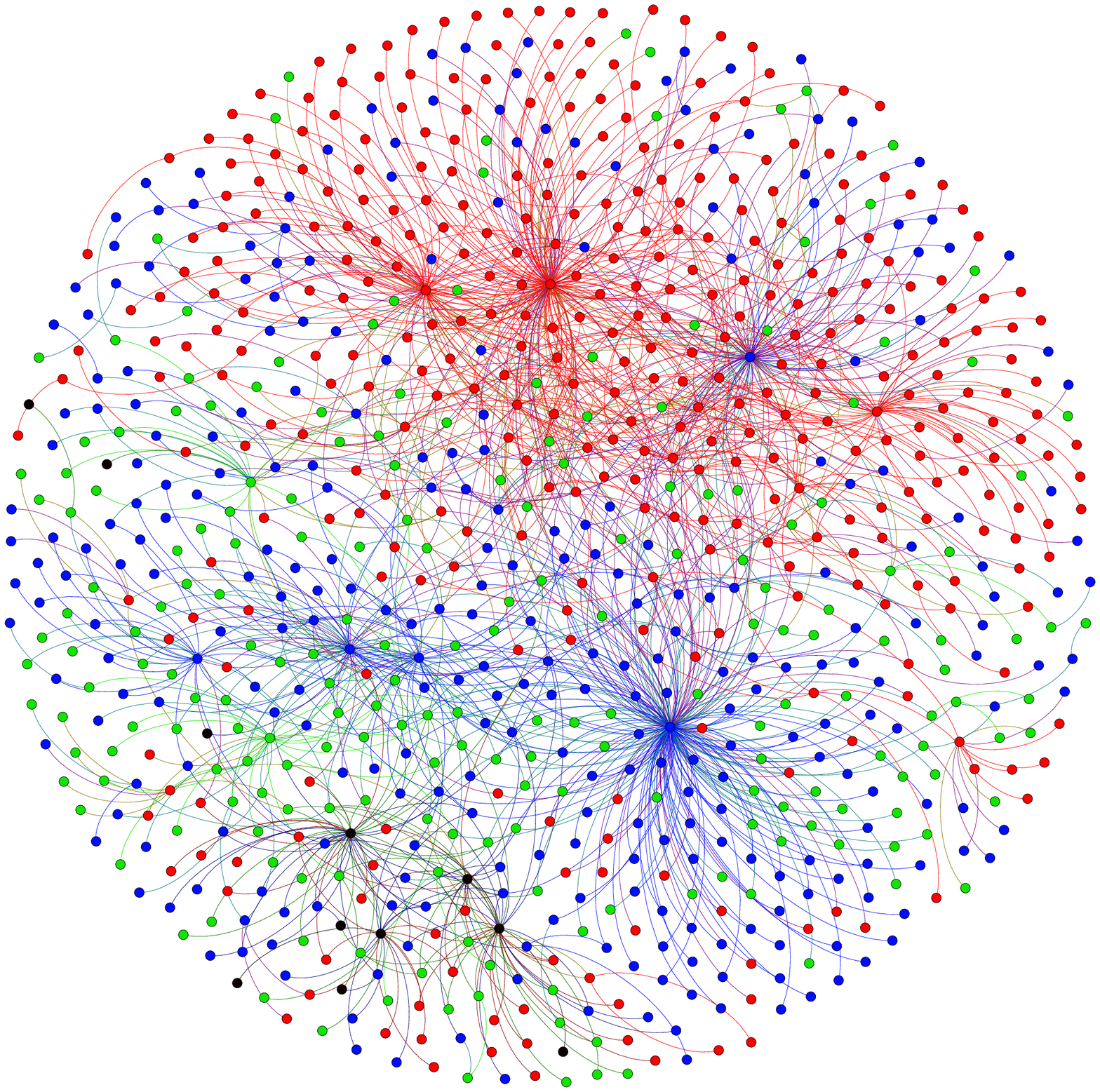}
 \label{fig:topo_emotion}}
\caption{The giant connected cluster of a network sample with $T=30.$ (a) is the network structure, in which each node stands for a user and the link between two users represents the interaction between them. Based on this topology, we color each node by its emotion, i.e., the sentiment with the maximum tweets published by this node in the sampling period. In (b), the red stands for \emph{anger}, the green represents \emph{joy}, the blue stands for \emph{sadness} and the black represents \emph{disgust}. The regions of same color indicate that closely connected nodes share the same sentiment.}
\label{fig:topology}
\end{figure*}

In this paper, the emotion is divided into four classes, including \emph{anger}, \emph{sadness}, \emph{joy} and \emph{disgust}. We employ the bayesian classifier developed in pervious work~\cite{moodlens}. In this method, we use the emoticon, which is pervasively used in Weibo, to label the sentiment of the tweets. At the first stage, 95 frequently used emoticons are manually labelled by different sentiments and then if a tweet only contains the emoticons of a certain sentiment, it would be labeled with this sentiment. From around 70 million tweets, 3.5 million tweets with valide emoticons are extracted and labeled. Using this data set as a training corpus, a simple but fast bayesian classifier is built in the second stage to mine the sentiment of the tweets without emoticons, which are about 95\% in Weibo. The averaged precision of this classifier is convincing and particularly the large amount of tweets we employ in the experiment can guarantee its accuracy further. Based on this framework, we demonstrate a sampled snapshot of interaction network with $T=30.$ As shown in Figure~\ref{fig:topo_emotion}, in which each user is colored by its emotion. We can roughly find that closely connected nodes generally share the same color, indicating emotion correlations in Weibo network. Besides, different colors show different clusterings. For example, the color of red, which represents \emph{anger}, shows more evident clustering. These preliminary findings inspire us that different emotions might have different correlations and a deep investigation is necessary.

\subsection{Emotion correlation}
\label{subsec:emotioncorrelation}

Emotion correlation is a metric to quantify the strength of sentiment influence between connected users. For a fixed $T$, we first extract an interaction network $G$ and all the tweets posted by the nodes in $G$. Then by employing the classifier established in the former section, the tweets for each user is divided into four categories, in which $f_1,$ $f_2,$ $f_3$ and $f_4$ represent the fraction of angry, joyful, sad and disgusting tweets, respectively. Hence we can use emotion vector $e_i(f_1^i,~f_2^i,~f_3^i,~f_4^i)$ to denote user $i$'s sentiment status. Based on this, we define pairwise sentiment correlation as follows. Given a certain hop distance $h$, we collect all user pairs with distance $h$ from $G$. For one of the four emotions $m(m=1,2,3,4)$ and a user pair $(j,q)$, we put the source user $j$'s $f_m^j$ into a sequence $S_m$, and the target user $q$'s $f_m^q$ to another sequence $T_m$. Then the pairwise correlation could be calculated by Pearson correlation as $$C_p^m=\frac{1}{l-1}\sum_{i=1}^{l}(\frac{S_i-\langle S_m\rangle}{\sigma_{S_m}})(\frac{T_i-\langle T_m
\rangle}{\sigma_{T_m}}),$$ where $\langle S_m\rangle=\frac{1}{l}\sum_{i=1}^{l}S_i$ is the mean, $\sigma_{S_m}=\sqrt{\frac{1}{l-1}\sum_{i=1}^{l}(S_i-\langle S_m\rangle)}$ is the standard deviation and $l$ is length of $S_m$ or $T_m$. Or it can also be obtained from Spearman correlation as $$C_s^m=1-\frac{6\sum_{i=1}^{l}d_i^2}{l(l^2-1)},$$ where $d_i$ is the rank difference between $S_i$ in $S_m$ and $T_i$ in $T_m.$ Intuitively larger $C_p^m$ and $C_s^m$ both suggest a more positive correlation for sentiment $m.$

Based on the dataset and classifier, interaction networks could be built and tweets of each user in the network would be emotionally labelled. Using the definition of correlations, we can then present the comparison of emotion correlations and the impact of local structures in the following section.

\section{Results}
\label{sec:results}

\begin{figure*}
\centering
 \subfloat[Person correlation]{\includegraphics[width=6cm]{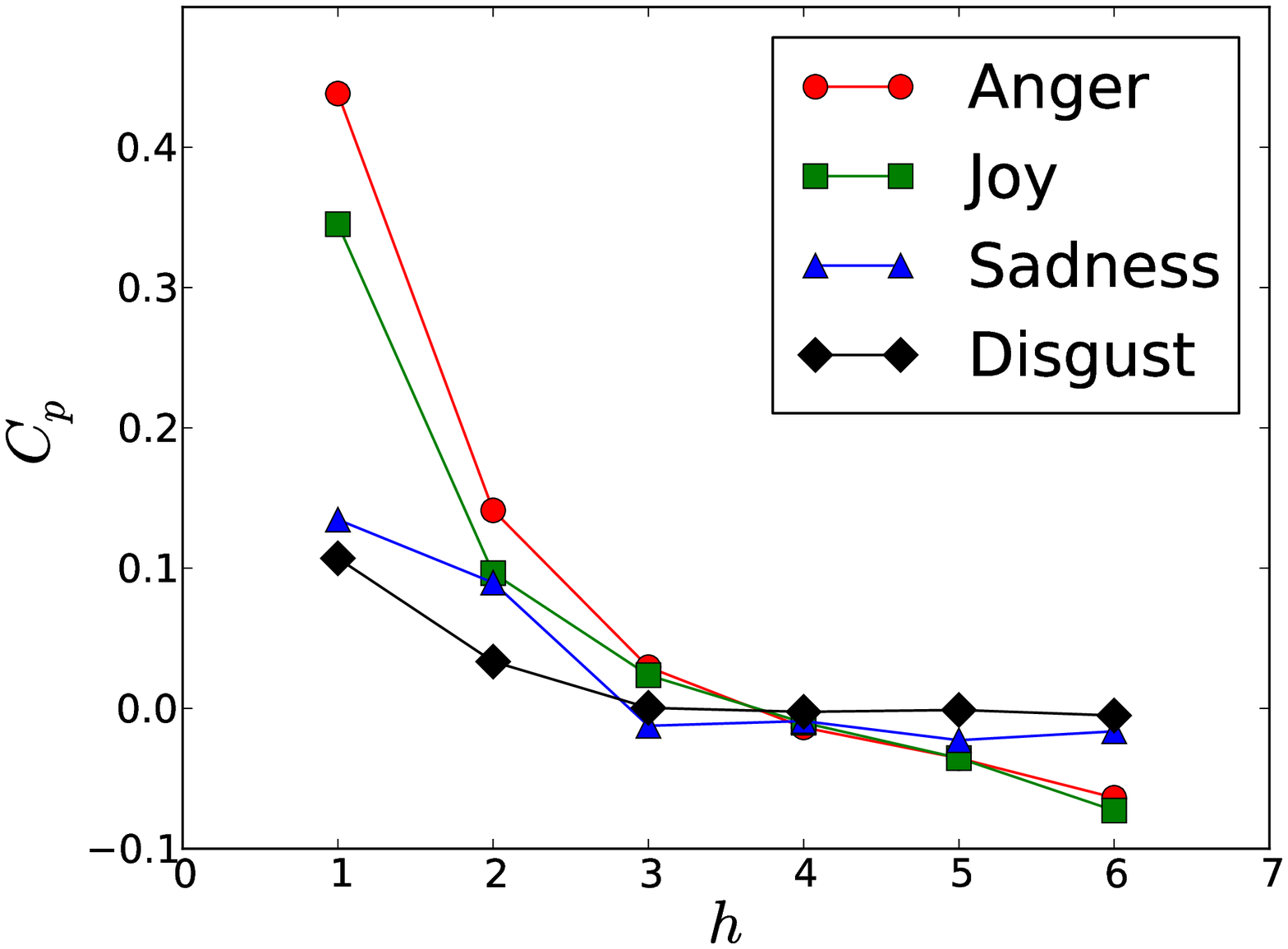}
 \label{fig:step_cp}}
 \subfloat[Spearman correlation]{\includegraphics[width=6cm]{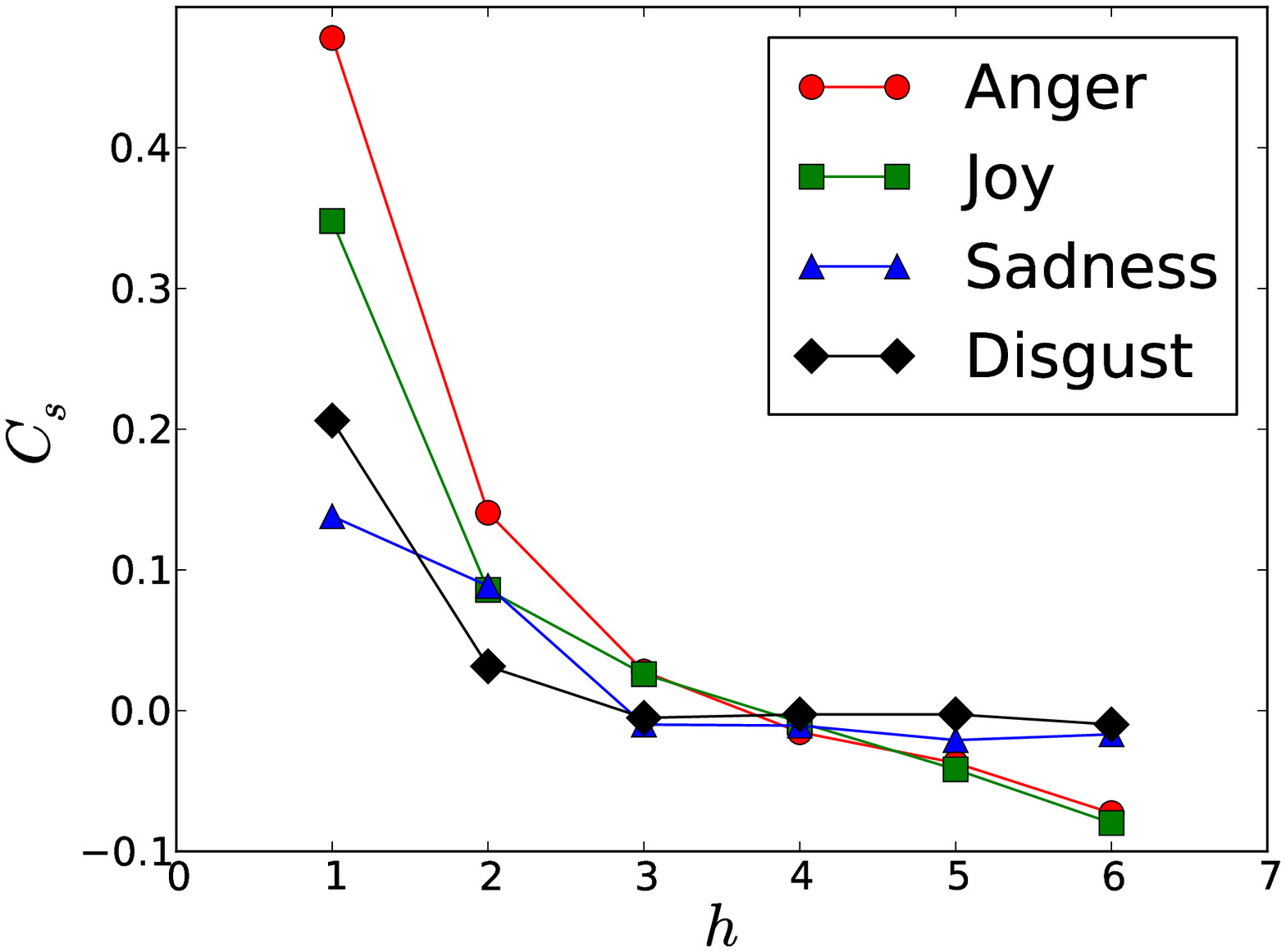}
 \label{fig:step_cs}}
\caption{Correlation for different emotions as the hot distance varies. Large $h$ means a pair of users are far away from each other in the social network we build. Here $T=30$ is fixed.}
\label{fig:step_c}
\end{figure*}

First we compare the correlation of different emotions based on the graph of $T=30,$ which ensures enough number of ties and users, and at the same time guarantees relatively high social tie strength. As shown in Figure~\ref{fig:step_c}, both Pearson correlation and Spearman correlation indicate that different sentiments have different correlations and \emph{anger} has a surprisingly higher correlation than other emotions. This suggests that \emph{anger} could spread quickly and broadly across the network because of its strong influence to the neighborhoods in the scope of about three hops. Although the previous studies~\cite{happiness_assortative,happiness_correlation} show that happiness is assortative in online social networks, but Figure~\ref{fig:step_c} further demonstrates that the correlation of \emph{anger} is much stronger than that of happiness. It means the information carrying angry message might propagate very fast in the network and this phenomenon is contrary to our intuition. While for \emph{sadness} and \emph{disgust}, they both share an unexpected low correlation even for small $h.$ For instance, the correlation of \emph{sadness} is less than 0.15 as $h=1$, which means sad status almost does not affect the directly connected friends at all. The results are also consistent with the previous findings that strength of the emotion correlation decreases as $h$ grows, especially after $h>6$~\cite{happiness_correlation}. In fact, as $h>3$, the emotion correlation becomes weak for all the sentiments, which means that the influence of the sentiment in the social network is limited significantly by the social distance. For example, for strong assortative emotions like \emph{anger} and \emph{joy}, their correlations just fluctuate around 0 as $h>5.$

\begin{figure*}
\centering
\subfloat[Person correlation]{\includegraphics[width=6cm]{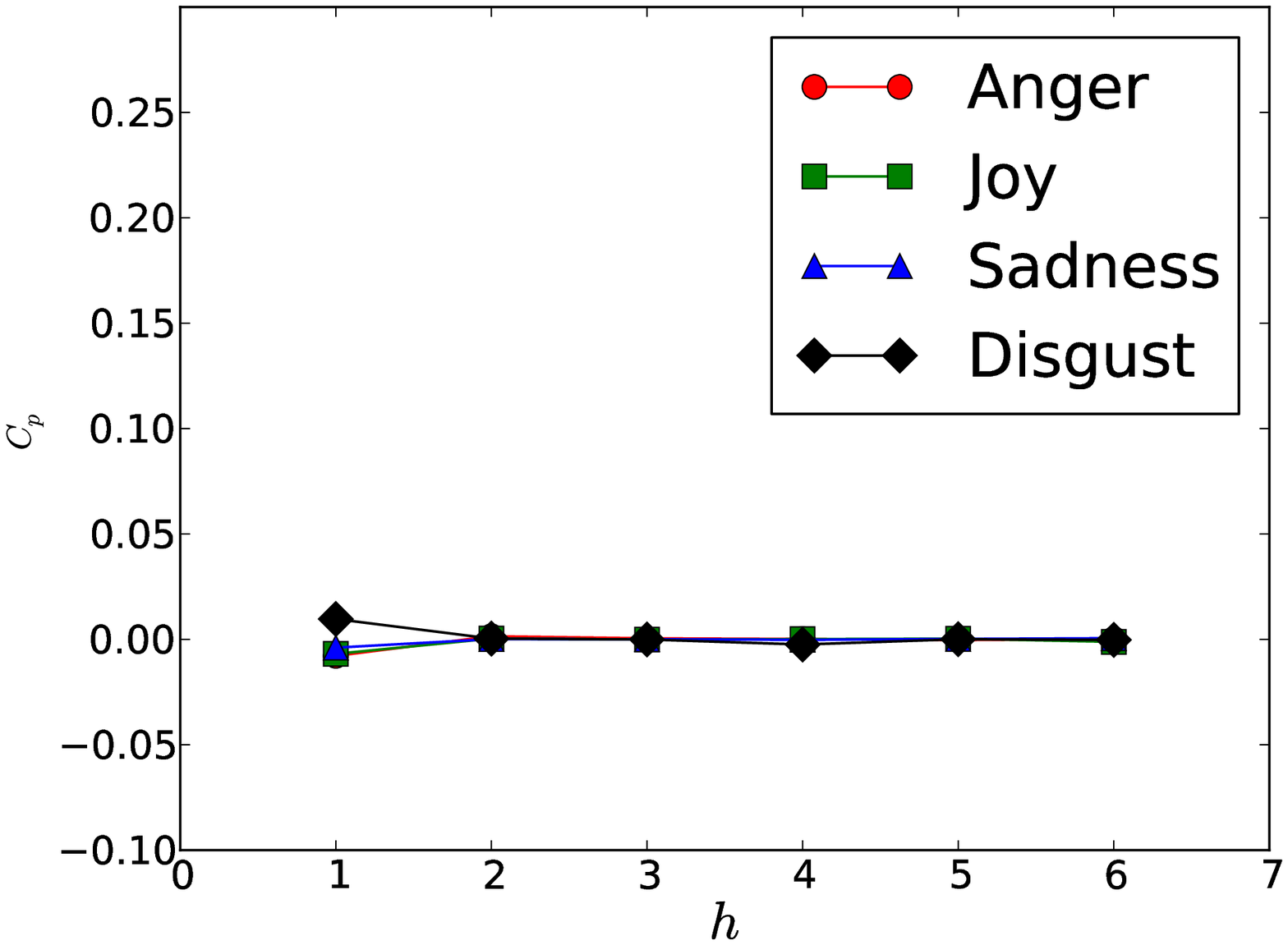}
\label{fig:step_shuffle_cp}}
\subfloat[Spearman correlation]{\includegraphics[width=6cm]{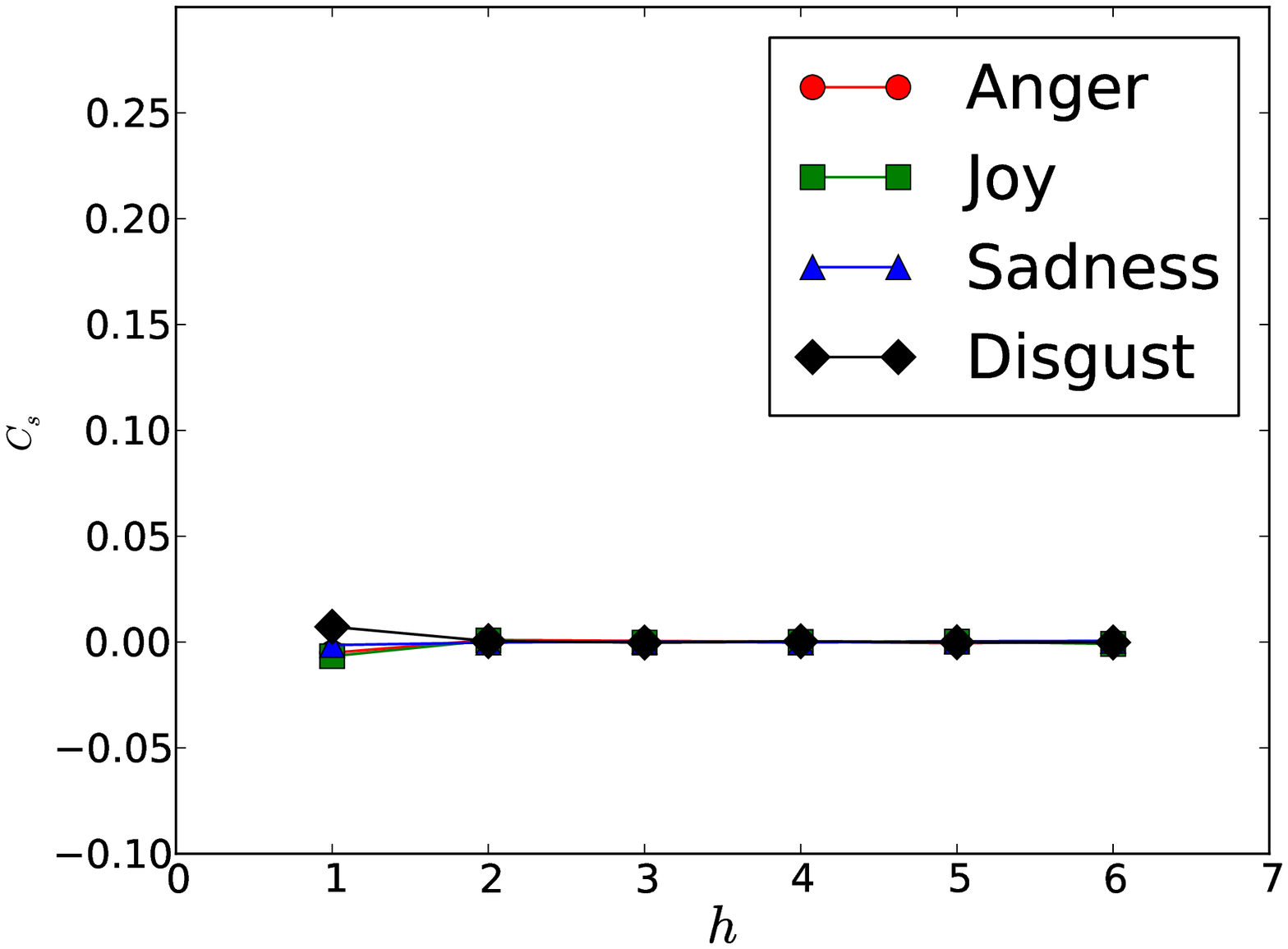}
\label{fig:step_shuffle_cs}}
\caption{The emotion sequence is randomly shuffled to test the correlation significance.}
\label{fig:step_shuffle}
\end{figure*}

In order to test the above correlation further, we also shuffle $S_m$ and $T_m$ randomly for sentiment $m$ and recalculate its correlation. As shown in Figure~\ref{fig:step_shuffle}, for the shuffled emotion sequence, there is no correlation existing for all the sentiments. It indicates that the former correlation we get is truly significant and for random pair of users in social network, there is no \emph{emotion homophily}. It further justifies that through social ties, the sentiment indeed spreads between closely connected friends and different users could influence their neighborhoods' mood statuses because of the social bonds between them. 

\begin{figure*}
\centering
\subfloat[\emph{anger}]{\includegraphics[width=5cm]{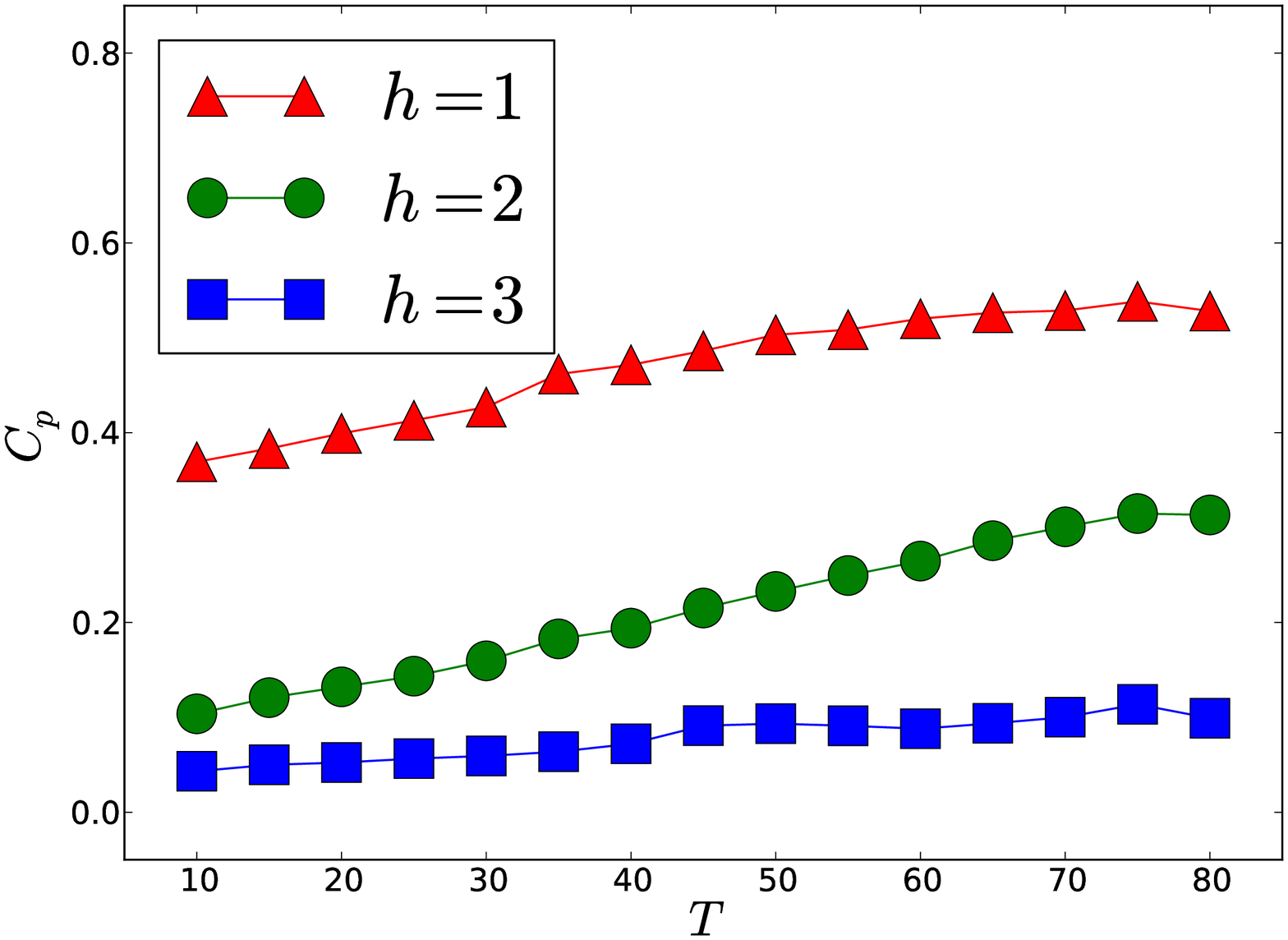}
\label{fig:T_anger_cp}}
\subfloat[\emph{joy}]{\includegraphics[width=5cm]{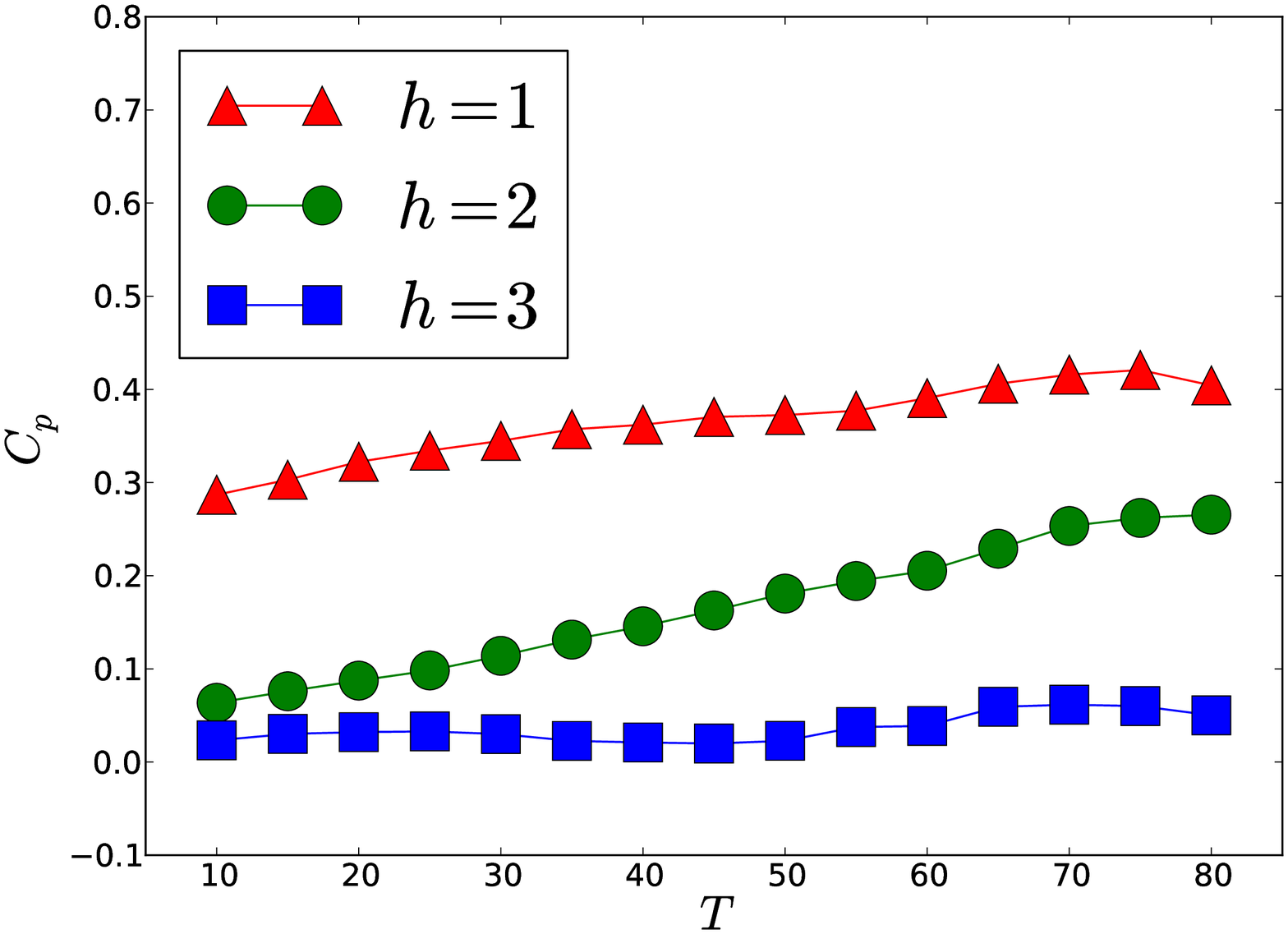}
\label{fig:T_joy_cp}}\\
\subfloat[\emph{sadness}]{\includegraphics[width=5cm]{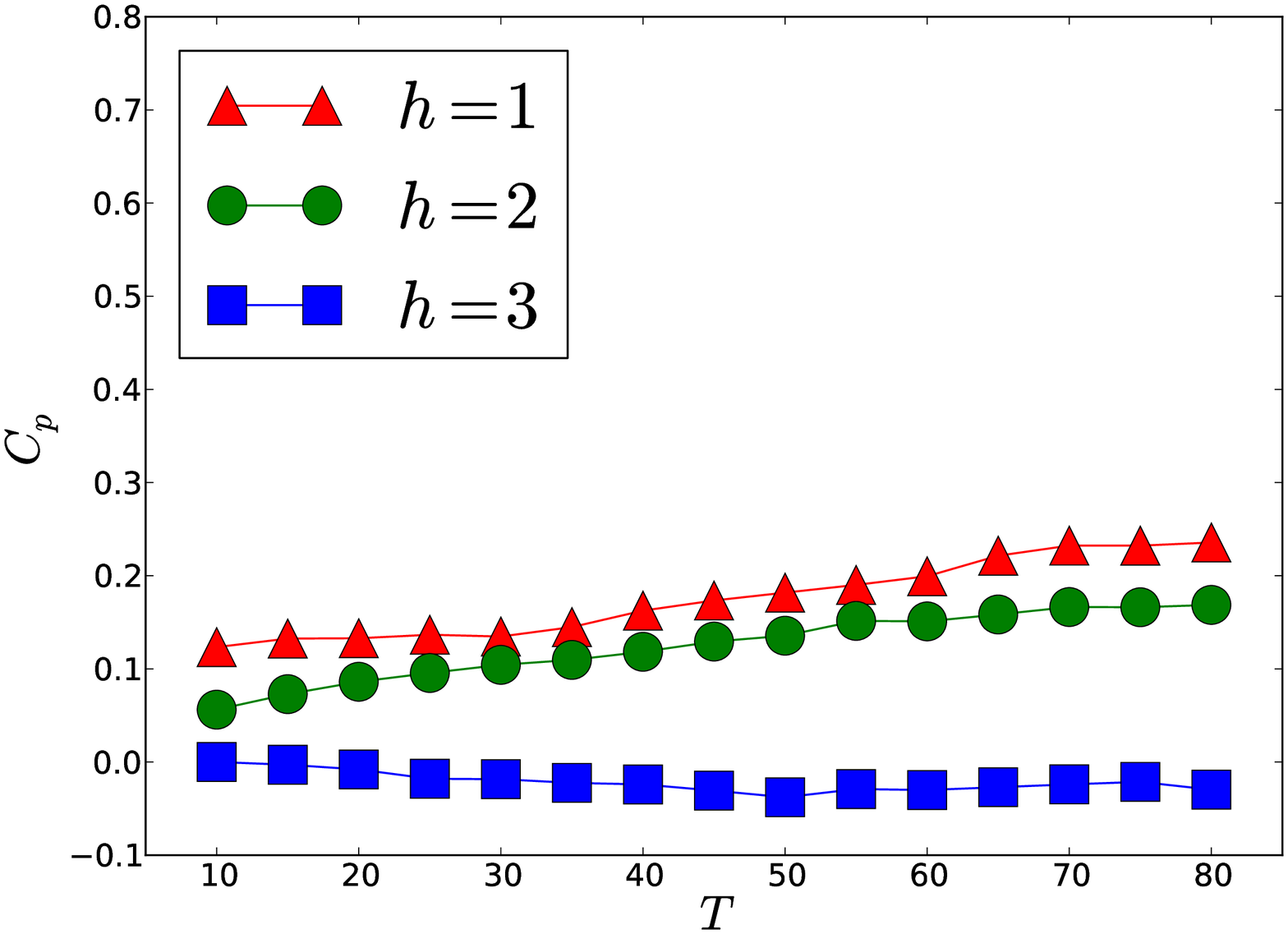}
\label{fig:T_sad_cp}}
\subfloat[\emph{disgust}]{\includegraphics[width=5cm]{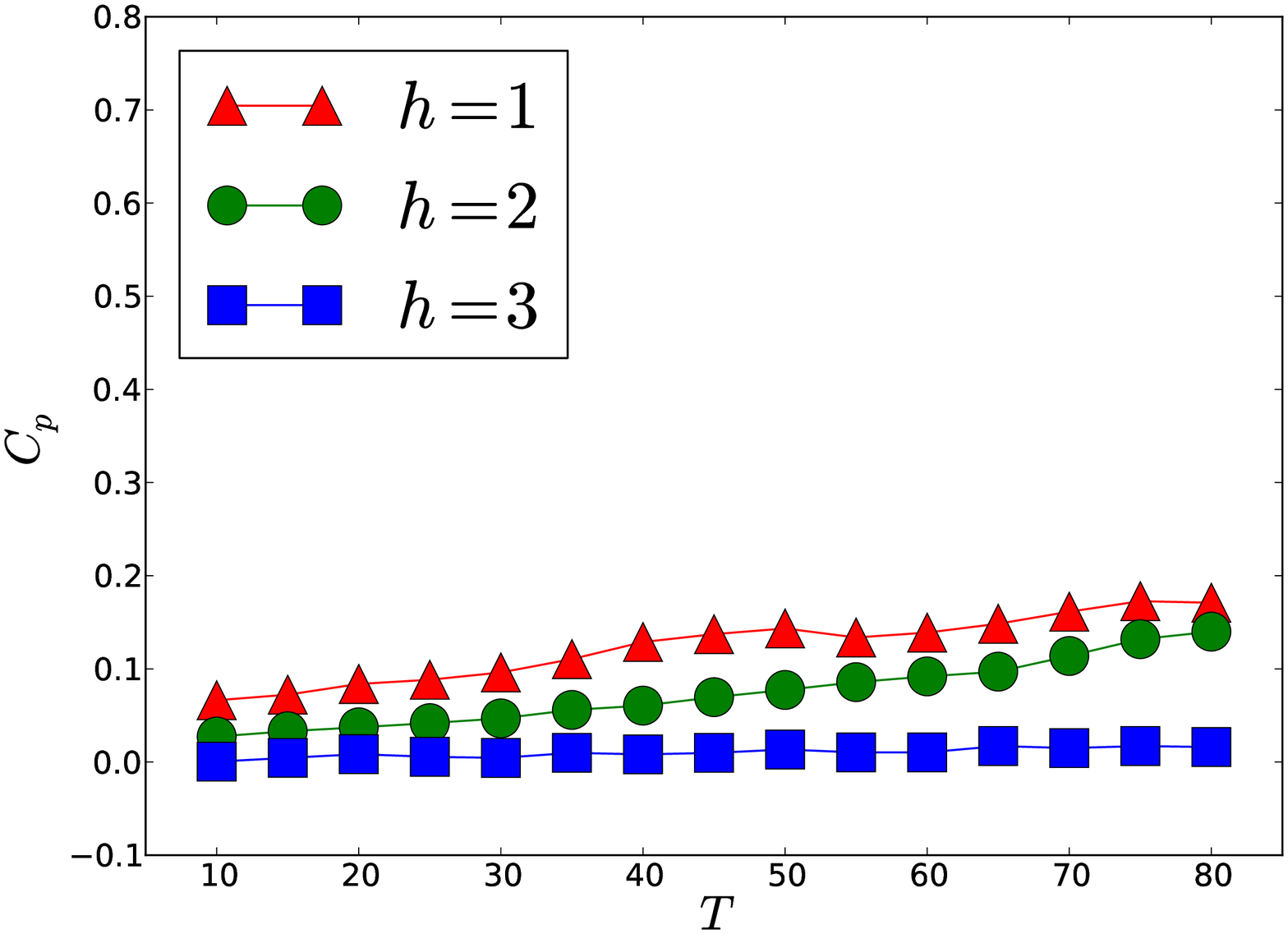}
\label{fig:T_disgust_cp}}
\caption{Pearson correlations of different $h$ for different networks extracted by varying $T.$ The case of $h>3$ is not considered here because of the weak sentiment correlation found in Figur~\ref{fig:step_c}.}
\label{fig:T_cp}
\end{figure*}

\begin{figure}
\centering
\includegraphics[width=7.5cm]{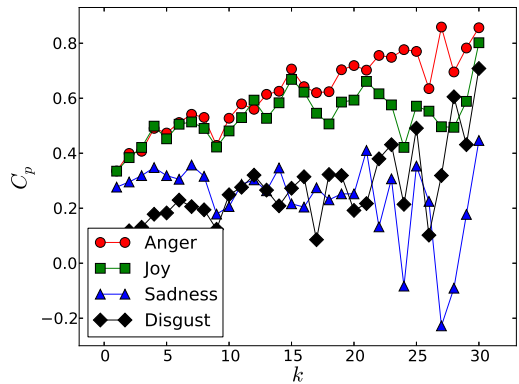}
\caption{Here $T$ is fixed to 30. Because the network is relatively small, the largest degree we get is only 30. The results just indicate when the degree is small, all sentiments' correlations increase with node degrees.}
\label{fig:k_cp}
\end{figure}

Investigating to what extent the local structure, like tie strength and node degree, could affect the emotion correlation is of importance for modeling sentiment influence and propagation. As shown in Figure~\ref{fig:T_cp}, we first disclose how the interaction threshold $T$ affects the sentiment influence. As discussed in Section~\ref{subsec:datasets}, larger $T$ produces smaller networks but with closer social relations and strong online interactions. It is also intuitive that frequent interactions in online social networks are positively related with strong social ties and convincing social bonds. Because of this, we can see in Figure~\ref{fig:T_cp} that for all the four emotions, their correlations inside two hops continue a steady increasing trend with $T$'s growth. Particularly for \emph{anger}, its Pearson correlation could rise to around 0.52. For weakly correlated emotions like \emph{sadness} and \emph{disgust}, although the correlation shows a slow growth for $h=1$ and $h=2$, while the maximum value of the correlation is still lower than 0.25. As $h=3$, the increment of the sentiment influence is trivial, especially for \emph{sadness} and \emph{disgust}. It illustrates that the primary factor of controlling emotion correlation is still the social distance and the social tie strength just functions for close neighbors in the scope of two hops. Secondly, we check the effect of users' degrees to the sentiment influence. We select a node $i$ with degree $k$ and then average its neighbors' emotion vectors to $e_i^{nei}(\frac{1}{k}\sum_jf_1^j,~\frac{1}{k}\sum_jf_2^j,~\frac{1}{k}\sum_jf_3^j,~\frac{1}{k}\sum_jf_4^j),$ where $j$ is an arbitrary neighbor of $i.$ Through adding $f_m^i$ into $S_m$ and $\frac{1}{k}\sum_jf_m^j$ into $T_m$, we could get the correlation of sentiment $m$ for the users with degree $k.$ As can be seen in Figure~\ref{fig:k_cp}, the sentiment correlation grows with $k$, especially for \emph{anger} and \emph{joy}, which illustrates that nodes with higher degrees in online social network posses more significant emotional influence to their neighborhoods. This finding is consistent with the conventional viewpoint that high degree nodes in the social network own more social influence and social capital. Specifically, the correlation of \emph{anger} and \emph{joy} are almost same for very small degrees, but later \emph{anger} shows a significant jump for large degrees and enlarges the gap as compared to \emph{joy}. As $k$ rieses to 30, the corrleation of \emph{anger} grows to 0.85. While the correlation of \emph{sadness} and \emph{disgust} do not demonstrate an obvious increasing trend and just fluctuate around 0.2 or even lower. It is worthy emphasizing that because the network size is small and we only have maximum degree around 30, which is far below the Dunbar's number～\cite{dunbarnumber}. We suspect that the correlation might stop rising if the degree is larger than Dunbar's number. The results of Spearman correlation are similar and not reported here.

To sum up, different emotions have different correlations in the social media. Compared to other sentiments, \emph{anger} has the most positive correlation, which indicates its fast and broad propagation. Local structure can affect the sentiment influence in near neighborhoods, from which we can learn that tie strength and node degree both could enhance the sentiment influence, especially for \emph{anger} and \emph{joy}, and their contributions to \emph{sadness} and \emph{disgust} are greatly limited. While high correlation of angry mood but weak influence of sad status indeed require much more detailed explorations to disclose the underlying reasons.

\section{Empirical Explanation}
\label{sec:explanation}

\begin{figure*}
\centering
 \subfloat[\emph{anger} phrases]{\includegraphics[width=6cm]{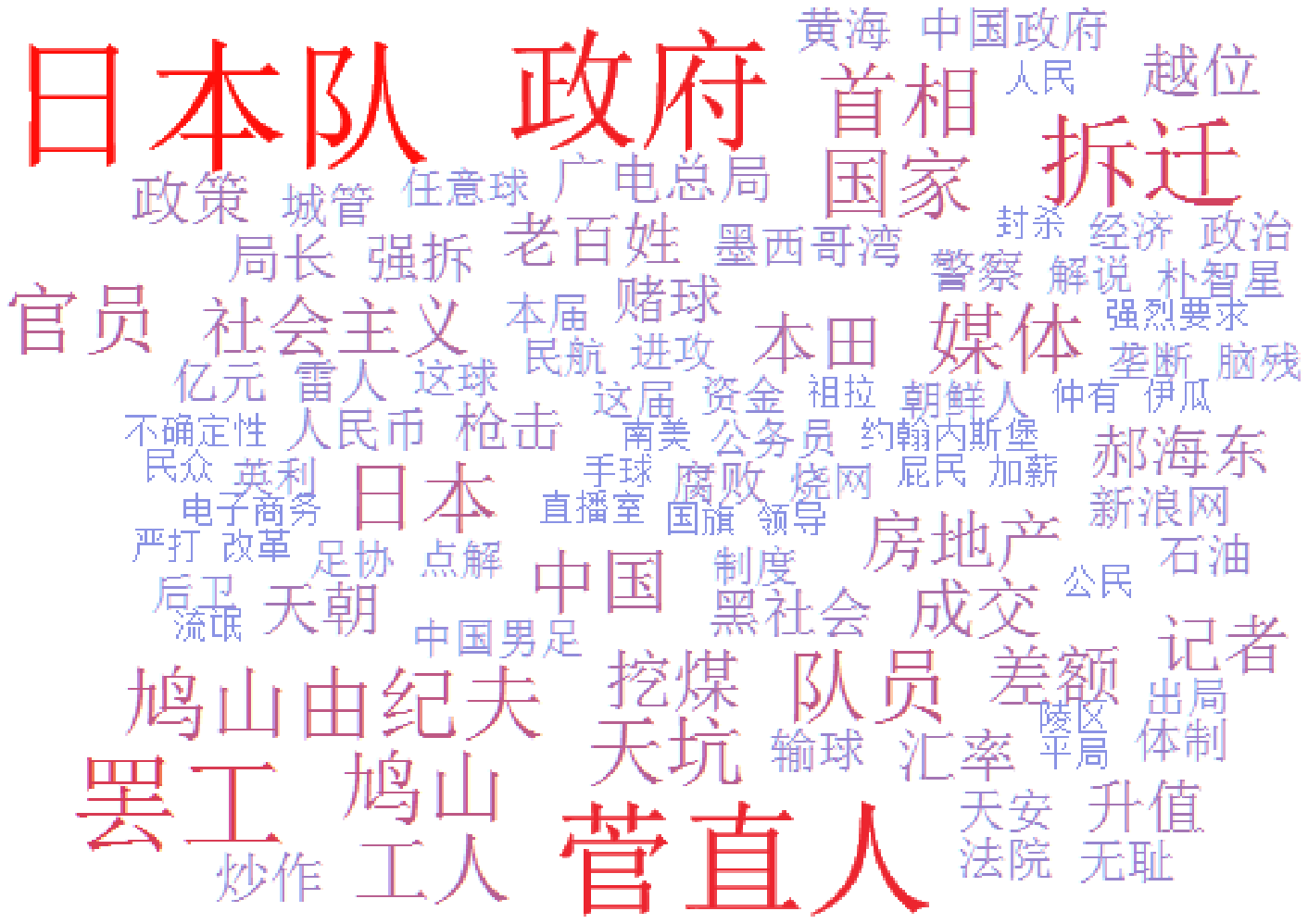}
 \label{fig:anger_keywords}}
 \subfloat[\emph{sadness} phrases]{\includegraphics[width=6cm]{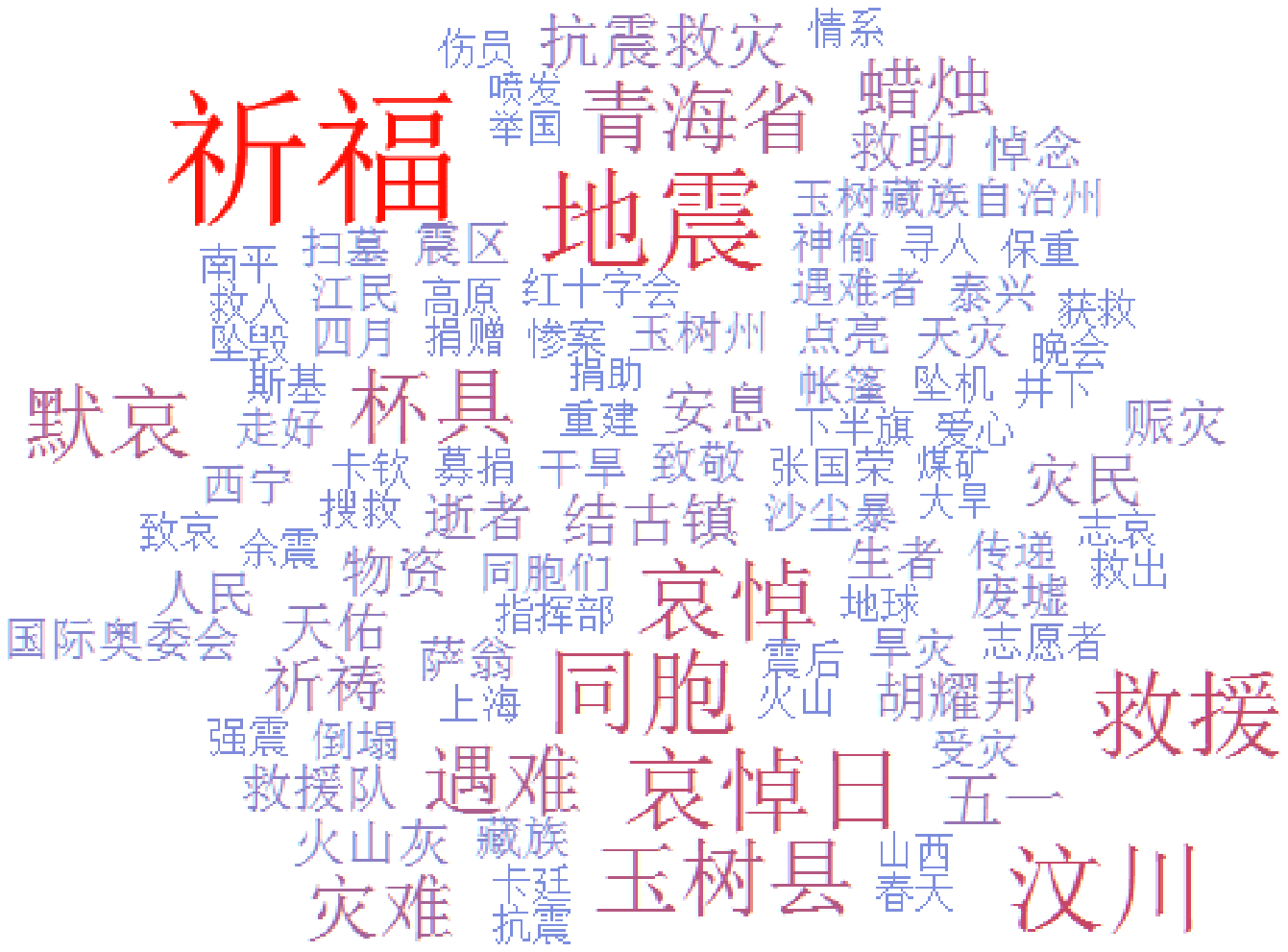}
 \label{fig:sadness_keywords}}
\caption{The example Chinese keywords extracted for \emph{anger} and \emph{sadness}, respectively. The top 20 keywords are also translated into English, which could be found through http://goo.gl/7JIEgR.}
\label{fig:keywords}
\end{figure*}

With the continuous growth, online social medias in China like Weibo have been becoming the primary channel of information exchange. In Weibo, the messages do not only deliver the factual information but also propagate the users' opinions about the social event or individual affairs. Hence we try to unravel the underlying reason of why \emph{anger} has a surprisingly high correlation but the spread of \emph{sadness} is weak from the view of keywords the corresponding tweets present. For a certain emotion $m$, we collect all the retweeted tweets(usually contain phrase like ``@'' or ``retweet'') with this sentiment in a specified time period to combine into a long text document. Focusing only on retweeted tweets could help reduce the impact of external media and just consider social influence from the social ties in Weibo. Several typical techniques are employed to mine the keywords or topic phrases from the documents, which are reported in Figure~\ref{fig:keywords}. Based on the keywords or topics we find, the real-world events or social issues could be summarized to understand the sentiment influence in detail.

With respect to \emph{anger}, we find two kinds of social events are apt to trigger the angry mood of users in Weibo. First one is the domestic social problems like food security, government bribery and demolition for resettlement. The ``shrimp washing powder'' which results in muscle degeneration and the self-burning event in Fenggang Yihuang County of Jiangxi province represent this category.  These events reflect that people living in China are dissatisfied about some aspects of the current society and this type of event can spread quickly as the users want to show their sympathy to the victims by retweeting tweets and criticizing the criminals or the government. Frequently appearing phrases like ``government'', ``bribery'', ``demolition'' and so on are strongly related with these events. The second type is about the diplomatic issues, such as the conflict between China and foreign countries. For instances, in August 2010, United States and South Korea held a drill on the Yellow Sea, which locates in the east of China. In September 2010, the ship collision of China and Japan also made users in Weibo extremely rageful. Actually, these events could arouse patriotism and stimulate the angry mood. Keywords like ``Diaoyu Island'', ``ship collision'' and ``Philippines'' show the popularity of these events at that time. To sum up, Weibo is a convenient and ubiquitously channel for Chinese to share their concern about the continuous social problems and diplomatic issues. Pushed by the real-world events, these users tend to retweet tweets, express their anger and hope to get resonance from neighborhoods in online social network. While regarding to \emph{sadness}, we find its strength of correlation is strongly affected by the real-world natural disasters like earthquake, as shown in Figure~\ref{fig:sadness_keywords}. Because the natural disaster happens occasionally and then the averaged correlation of the sadness is very low and the strength of its correlation might be highly fluctuated.

In summary, real-world society issues are easy to get attention from the public and people tend to express their anger towards theses issues through posting and retweeting tweets in online social media. The angry mood delivered through social ties could boost the spread of the corresponding news and speed up the formation of public opinion and collective behavior. This can explain why the events related to social problems propagate extremely fast in Weibo.

\section{Discussion and Conclusion}
\label{sec:conclusion}

Users with similar demographics have high probabilities to get connected in both online and offline social networks. Recent studies reveal that even the psychological states like happiness are assortative, which means the happiness or well-being is strongly correlated between connected users in online social medias like Twitter. Considering the oversimplification of the sentiment classification in the previous literature, we divide the emotion into four categories and discuss their different correlations in details based on the tweets collected from Weibo of China, and the dataset has been public available to research community.

Our results show that \emph{anger} is more influential than other emotions like \emph{joy}, which indicates that the angry tweets can spread quickly and broadly in the network. While out of our expectation, the correlation of \emph{sadness} is low. Through keywords and topics mining in retweeted angry tweets, we find the public opinion towards social problems and diplomatic issues are always angry and this extreme mental status also boost the propagation of the information in Weibo. This might be the origin of large scale online collective behavior in Weibo about society problems such as food security and demolition for resettlement in recent years. We conjecture that \emph{anger} plays a non-ignorable role in massive propagations of the negative news about the society, which are always hot trends in today's internet of China.

Besides, we also investigate the affect of local structure to the emotion correlation in online social media, which is not fully probed before. We find that for a pair of users the emotion correlation is stronger if more interactions happen between them. We also disclose that the a node's degree could significantly enhance the sentiment influence to its neighborhood, especially for \emph{anger} and \emph{joy}. These findings could shed light on modeling sentiment influence and spread in social networks.

\section{Acknowledgements}
\label{sec:ack}

This work was partially supported by the fund of the State Key Laboratory of Software Development Environment under Grant SKLSDE-2011ZX-02, the Research Fund for the Doctoral Program of Higher Education of China under Grant 20111102110019, and the National 863 Program under Grant 2012AA011005. JZ and YC both thank the Innovation Foundation of BUAA for PhD Graduates.





\bibliographystyle{elsarticle-num}
\bibliography{refs}







\end{document}